\input{aipcheck}

\documentclass[bibliocites
    ,final            
  ]
  {aipproc}

\layoutstyle{6x9}

\newcommand{\rem}[1]{ }

\newcommand{\beq}{\begin{equation}}
\newcommand{\eeq}{\end{equation}}

\begin{document}

\title{Radiation Processes in GRBs. Prompt Emission}

\author{Mikhail V. Medvedev}{
  address={Department of Physics and Astronomy, University of Kansas,
Lawrence, KS 66045}
}



\begin{abstract}
A substantial fraction of prompt GRB spectra have soft spectral indexes
exceeding the maximum allowed by the synchrotron model $\alpha_{max}=-2/3$. 
Some spectra also exhibit very sharp break at $E_p$, inconsistent with the 
smooth synchrotron spectra. These facts pose a serious problem for the 
``optically thin synchrotron'' interpretation of the prompt emission.
We review various models suggested in order to resolve this puzzle.  
\end{abstract}

\maketitle


\section{Introduction}

Time-resolved spectral analyzes of {\em BATSE} and {\em BeppoSAX}
\citep{Preece+00,Frontera+00} clearly demonstrate that 30-50\%
of spectra violate the so-called ``synchrotron line of death'' (LoD), 
i.e., they have the soft photon indexes $\alpha$ greater than $-2/3$
(note, $F_\nu\propto\nu^{\alpha+1}$).
In addition, a significant number of the spectra are better fit
with the sharply broken power-law (BPL) model than with the smooth Band 
function. These facts make the simplest synchrotron interpretation of 
the prompt GRB emission at least questionable. Some attention has
been paid to this problem and here we review alternative models 
suggested by several authors.

\section{Self-absorbed synchrotron model}

\begin{figure}
  \includegraphics[height=.2\textheight]{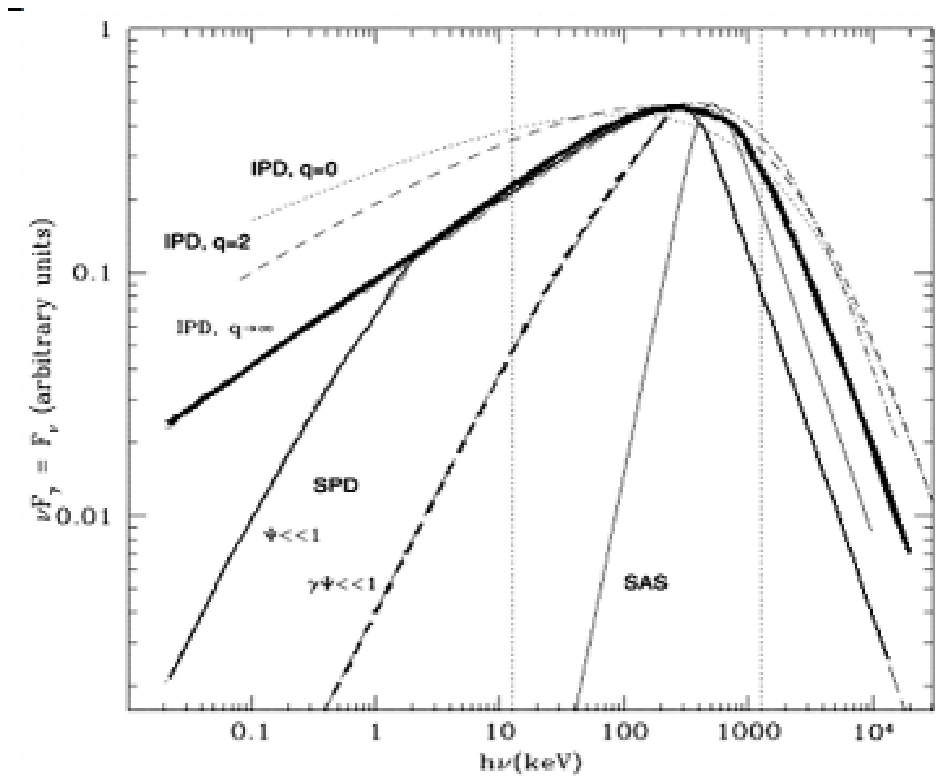}
\hskip2em
  \includegraphics[height=.2\textheight]{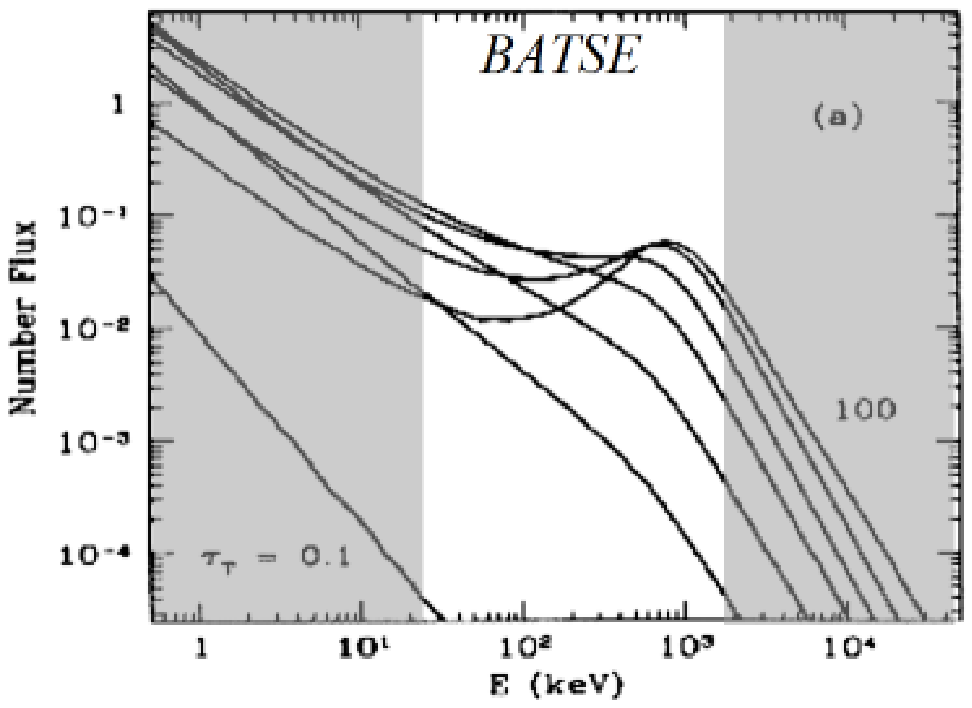}
  \caption{{\it (a)} The synchrotron self-absorbed spectrum (SAS) and 
the synchrotron spectrum in the small pitch-angle regime (from \cite{LP02}),
the thick solid line shows a standard synchrotron spectrum for comparison.
{\it (b)} The spectrum produced by saturated Comptonization 
(from\cite{Liang+97}).
\label{SAS+SC}}
\end{figure}

The simplest model which can produce a hard spectrum at low energies
suggests that synchrotron radiation may be self-absorbed. This possibility
has been considered by several authors; for more discussion and
references, see Ref. \cite{LP02}. The low-energy power-law index depends on 
the relative values of the self-absorption  frequency $\nu_a$ and the
peak synchrotron frequency $\nu_m$:
\begin{equation}
F_\nu\propto\left\{
\begin{array}{ll}
\nu^{5/2}, & \textrm{ for } \nu_m<\nu\ll\nu_a; \\ 
\nu^{1/3}, & \textrm{ for } \nu_a<\nu<\nu_m; \\ 
\nu^{2}, & \textrm{ for }\nu\ll\textrm{ min}(\nu_m,\nu_a).
\end{array}
\right.
\end{equation}
Note that the second case corresponds to the optically thin regime. 
A typical self-absorbed spectrum with $\nu_m<\nu_a$ is shown in 
Fig. \ref{SAS+SC}(a) by the curve labeled SAS. In addition to a large 
spectral index, a self-absorbed spectrum has also a much narrower peak
than the optically thin spectrum. Both these properties often lead to
improved spectral fits of LoD-violating and BPL bursts. 

What conditions of a fireball are needed to have the self-absorption 
frequency in the {\em BATSE}'s spectral window? 
The optical depth to synchrotron self-absorption
\begin{equation}
\tau\sim\left(\frac{l}{10^{13}\textrm{ cm}}\right)
\left(\frac{n}{10^8\textrm{ cm}^{-3}}\right)
\left(\frac{B}{10^8\textrm{ G}}\right)^{2/3}
\left(\frac{\gamma_{m}}{50}\right)^{-8/3}
\left(\frac{\Gamma}{10^3}\right)^3
\left(\frac{\nu_{obs}}{10^{19}\textrm{ Hz}}\right)^{-5/3}
\end{equation}
must be of order unity for the observed frequency $\nu_{obs}$ to be in the
{\em BATSE} range. Here $l$ and $n$ are the line-of-sight path 
length and particle density in the co-moving frame, $\gamma_m$ and $\Gamma$
are the minimum Lorentz factor of power-law electrons and the bulk Lorentz 
factor of the ejecta, and $B$ is the co-moving magnetic field strength.
Apparently, the values of the parameters are rather extreme, e.g., the 
magnetic field strength is (much) greater than the equipartition field of 
$\sim 10^5...10^6$~G, typically assumed within the standard synchrotron 
shock model (SSM). Another problem of this model is a very low efficiency 
of the fireball shock because the peak synchrotron frequency (where 
most of the energy is emitted) is deeply in the optically thick range. 

\section{Saturated Comptonization model}

Another model that may be of interest to us is the so-called
{\em saturated synchrotron self-Compton} proposed in a series 
of papers (see e.g., Ref. \cite{Liang+97}) primarily in an attempt
to explain the spectral peak energy -- fluence anti-correlation 
observed in several long, bright, smooth GRBs. The model proposes that
impulsively accelerated, non-thermal, relativistic electrons (and,
perhaps, pairs) repeatedly Compton up-scatter self-emitted radio/infrared
synchrotron photons into gamma-ray energies. The Thompson optical depth
is initially large, $\tau_T\gg1$, so that the emerging gamma-rays are in 
thermal equilibrium with electrons and $\alpha$ approaches the Wien limit,
$\alpha=+2$, whereas the synchrotron soft-photon source is strongly 
self-absorbed by internal free-free and synchrotron opacities. 
As time goes on, the Thompson opacity decreases and for $\tau_T\ll1$
the spectrum reduces to a single-scattering Compton spectrum with the slope 
$\alpha=-(p+1)/2$ (for the electron distribution $N(E)\propto E^{-p}$).
Thus, this model can naturally explain the hard-to-soft evolution
in prompt GRB spectra. 

Typical spectra are shown in Fig. \ref{SAS+SC}(b) for various values of 
$\tau_T$. It is quite clear that the emerging spectrum differ dramatically
from the Band spectrum (even within a narrow {\em BATSE} window) which
nicely fits the majority of GRBs. The required values of the fireball 
parameters are also not very likely: the comptonizing electrons ought to
be ``warm'' with $\gamma_e\sim$~few (in contrast to the SSM, in which
electrons are in near equipartition, hence $\gamma_e\sim1000$) and 
the required magnetic fields are also too week, $B\sim0.1...10$~G.

\section{Photospheric model}

\begin{figure}
  \includegraphics[height=.2\textheight]{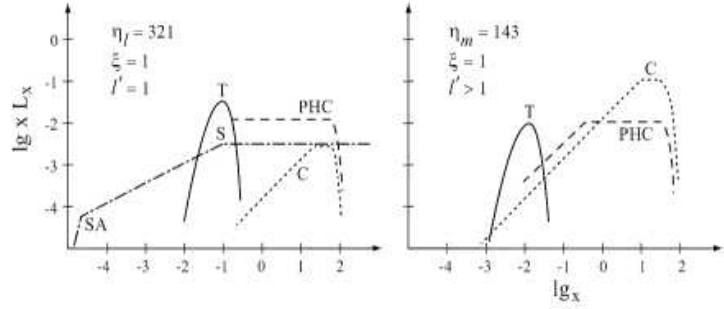}
  \caption{Spectra predicted by the photospheric model (from \cite{MR00}).
\label{P}}
\end{figure}

A synthetic model incorporating a standard synchrotron internal shock model 
and an extended photosphere can also explain steep low-energy spectra 
\cite{MR00}. In this model synchrotron photons are re-processed in the 
photosphere. The single-scattering Comptonized and photosphetic 
components naturally have low-energy spectra with $\alpha=0$. 
Fig. \ref{P} shows some examples; here T: thermal photosphere, 
PHC: photospheric comptonized component, S: shock synchrotron, 
C: shock paid-dominated comptonized component.
For a detailed discussion the reader is referred to Ref. \cite{MR00}.
The weakness of this model is that it requires moderate Thompson
opacities $\tau_T\sim1$ which, in turn, requires either fine tuning
of plasma parameters or, alternatively, some sort of self-regulated 
pair opacity which produces the column density which self-adjusts itself 
to a column density of few g cm$^-2$. This model also requires very low 
baryonic load, that is, large bulk Lorentz factors $\Gamma>1000$.

\section{Small pitch-angle radiation model}

Optically thin synchrotron radiation has a low-energy asymptotic power-law 
index $\alpha=-2/3$ only if the the emitting electrons have an isotropic
distribution. For electrons having anisotropic velocity distribution
this may not be the case. Let us consider a beam of mono-energetic 
highly-relativistic  electrons ($\gamma_e\gg1$ is their Lorentz factor) 
propagating almost along a homogeneous magnetic field, so that the 
parallel velocity is much larger than the transverse velocity:
\begin{equation}
v_\perp/v_\| \sim\Psi\ll 1/\gamma_e,
\end{equation}
where $\Psi$ is the pitch-angle. In this case, from $v^2=v_\|^2+v_\perp^2$
and $1-v^2=1/\gamma_e^2$ it follows that the transverse motion of 
such electrons is non-relativistic, $v_\perp\ll1$. Radiation emitted 
by these electrons will be {\em cyclotron} (not synchrotron), relativistically 
boosted with the Lorentz factor $\sim\gamma_e$ along the magnetic field.
The low-energy asymptotic spectrum of cyclotron radiation is steeper,
$\alpha=0$. This is called the ``small pitch-angle'' regime.

Small pitch-angle radiation has been suggested as yet another way of
producing steep low-energy spectra, see Ref. \cite{LP02} and references
therein. The typical spectra for $1/\gamma_e<\Psi\ll1$ and 
$\Psi\ll 1/\gamma_e$ are shown in Fig. \ref{SAS+SC}(a) by the curves labeled 
SPD. Note that in the latter case, the spectrum has a very sharp break. Thus, 
this model can naturally explain both LoD-violating and BPL bursts. The small 
pitch-angle radiation model relies, however, on a crucial assumption: a highly
anisotropic electron distribution is somehow created and maintained 
at the shock. Moreover, it is a well-known fact that (highly) anisotropic
particle distributions are always unstable with respect to a number
of plasma instabilities. Finally, in the small pitch-angle regime only
the transverse energy $\sim m_e v_\perp^2/2$ can be converted into
radiation. Because of the condition $v_\perp/v_\| \ll 1/\gamma_e$ with 
$\gamma_e\sim1000$ or more, the radiation efficiency will be enormously low.

\section{Jitter radiation model}

\begin{figure}
  \includegraphics[height=.26\textheight]{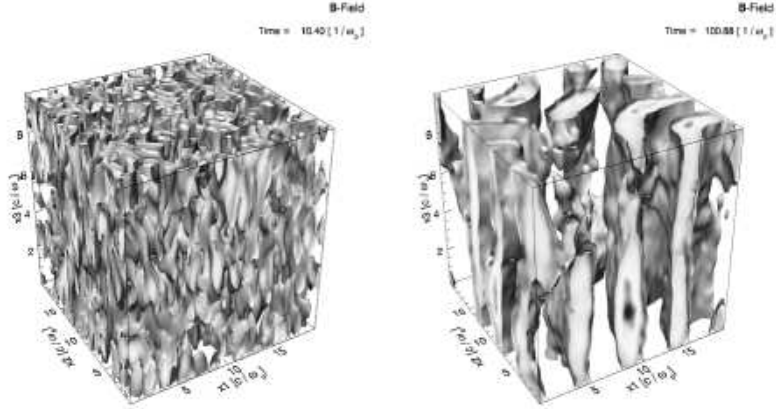}
  \caption{The magnetic field 3D structure (surfaces of constant $B^2$) 
at a relativistic shock (from \cite{Silva+03}). 
\label{3D}}
\end{figure}

It is now becoming a widely accepted fact that magnetic fields of
sub-equipartition strength are generated at the front of a 
relativistic shock via the two-stream (or Weibel) instability.
The magnetic field generation has been predicted theoretically 
\cite{ML99} and then confirmed via 3D PIC kinetic simulations
\cite{Silva+03,Fred.+03}. The produced magnetic fields have 
rather unusual properties. The field is predominantly generated in 
the direction, perpendicular to the shock propagation direction.
In the plane of the shock, the field is highly chaotic with the 
correlation length being of order the relativistic 
skin depth 
\begin{equation}
\lambda_s\simeq \frac{c\sqrt{\gamma_s}}{\omega_{ps}}
\sim (3\textrm{ cm})\;\gamma_e^{1/2}\left(\frac{m_s}{m_e}\right)^{1/2}
\left(\frac{n_s}{10^{10}\textrm { cm}^{-3}}\right)^{-1},
\end{equation}
where $\omega_{ps}=(4\pi e^2 n_s/m_s)^{1/2}$ is the non-relativistic
co-moving plasma frequency of species $s=e^-, p$ (both species, electrons 
and protons, generate the field). A typical strength of the field
is $\epsilon_B=B^2/(8\pi\Gamma n m_pc^2)\sim10^{-3}$ (here $n$ is the
density downstream). The correlation scale $\lambda$ is not
constant, rather it increases with time, i.e., with the distance
from the shock front. Fig. \ref{3D} represents three-dimensional 
contours of constant $B^2$ close to the shock front and far downstream.


\begin{figure}
  \includegraphics[height=.1\textheight]{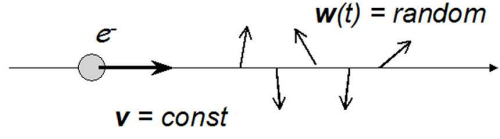}
  \caption{An electron motion in the jitter regime.
\label{motion}}
\end{figure}

\begin{figure}
  \includegraphics[height=.2\textheight]{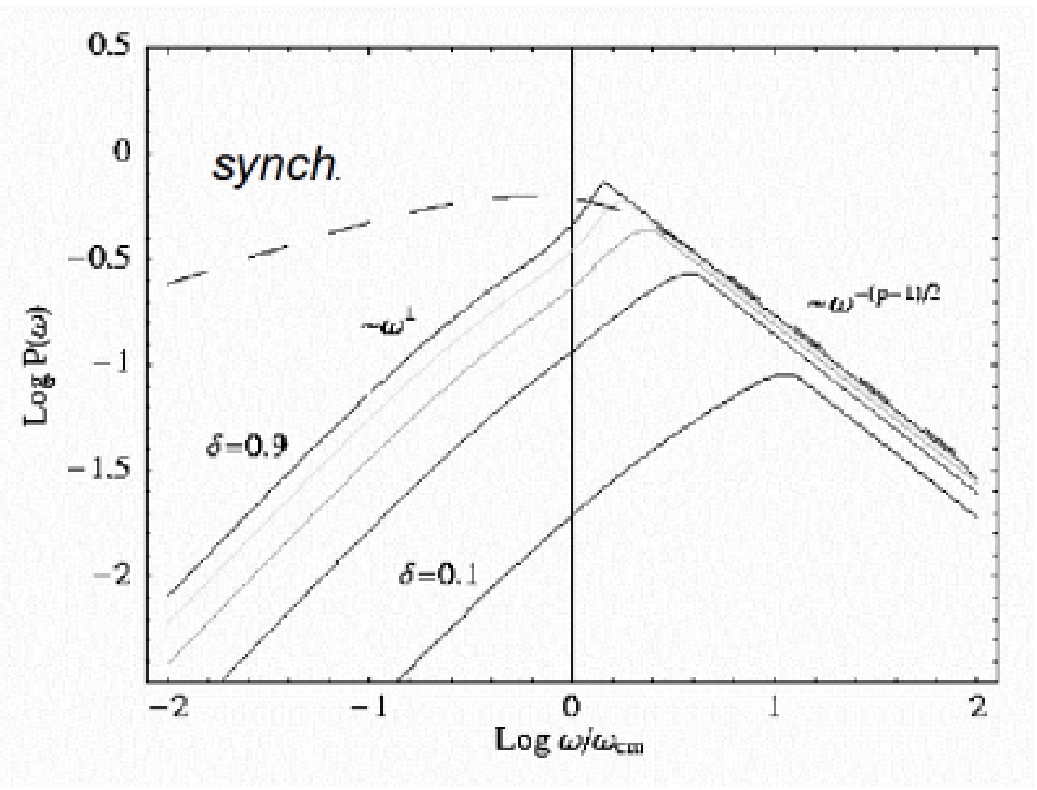}
\hskip2em
  \includegraphics[height=.2\textheight]{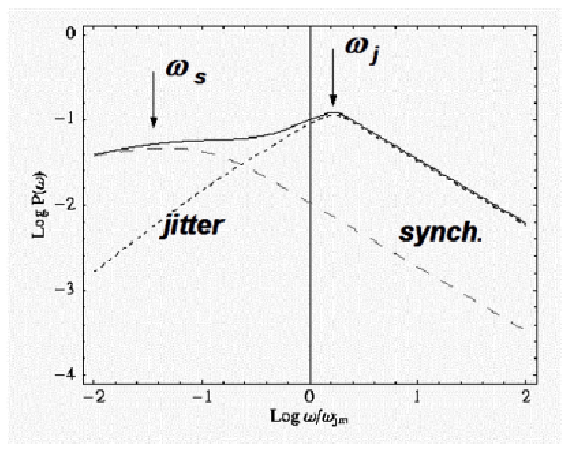}
  \caption{{\it (a)} The spectrum of jitter radiation for the power-law 
distributed electrons (from \cite{M00}). {\it (b)} A generalized 
jitter+synchrotron spectral model (from \cite{M00}).
\label{spec-j+s}}
\end{figure}

It can straightforwardly be evaluated that the correlation length of the
electron-produced field, $\lambda_e$, is much smaller than the Larmor
radius of a relativistic radiating electron, $\rho_e$. The electron 
trajectory is not helical, so the standard synchrotron theory is not 
applicable. Quantitatively, jitter regime (for details, see 
Ref. \cite{M00}) occurs when the deflection angle 
of the electron is smaller than the relativistic beaming angle 
$\sim1/\gamma_e$, i.e.:
\begin{equation}
\delta\sim{\lambda_e}/({\rho_e/\gamma_e})\sim({eB\lambda_e})/({m_ec^2})<1.
\end{equation}
In the case $\delta\ll1$, the particle motion may safely be approximated 
as straight. As the electron moves at a constant velocity, it experiences 
short accelerations in random directions, perpendicular to the direction 
of motion, as represented in Fig. \ref{motion}. The power spectrum of 
radiation  is obtained from the Lienard-Wichert potentials:
\begin{equation}
\frac{dW}{d\omega}=\frac{e^2\omega}{2\pi c^3}\int_{\omega/2\gamma_e^2}^\infty
\frac{\left|{\bf w}_{\omega'}\right|^2}{\omega'^2}
\left(1-\frac{\omega}{\omega'\gamma_e^2}+\frac{\omega^2}{2\omega'^2\gamma_e^4}
\right)\,d\omega',
\end{equation}
where ${\bf w}_{\omega'}=\int{\bf w}e^{i\omega't}\,dt$ is the Fourier 
component of the particle's acceleration, which is related to the
spectrum of the magnetic field as 
$w_{\omega'}=({eB_{\omega'}})/({\gamma_e m_e})=({eB_{k'}})/({\gamma_e m_ec})$.
For a standard energy distribution of electrons (power-law with a cutoff
at low energies, $\gamma_{e,min}$), the resultant spectrum is shown in 
Fig. \ref{spec-j+s}(a). It is well described by a BPL model with $\alpha=0$, 
the high energy exponent $\beta=-(p+1)/2$, and the jitter break frequency:
\begin{equation}
\nu_j\simeq(c/\lambda_e)\;\gamma_{e,min}^2\;\Gamma
\sim(10^{10}\textrm{ Hz})\;\gamma_{e,min}^{3/2}\;\Gamma.
\end{equation}
Note that this frequency is independent of the magnetic field strength.

Unlike the electron-produced fields, the proton-produced magnetic field
has a larger spatial correlation scale, $\lambda_p$, for which $\delta>1$.
An electron radiates synchrotron radiation in such a field. Also, a 
large-scale magnetic field may be ejected from a magnetized progenitor.
Therefore, in general, the spectrum may consist of two components,
a jitter component (due to small-scale fields, $B_{SS}$, with $\delta<1$)
and a synchrotron component (due to large-scale fields, $B_{LS}$, 
with $\delta>1$), as in Fig. \ref{spec-j+s}(b). 
The jitter-to-synchrotron peak frequency ratio and the ratio of the 
{\em photon} fluxes at these peak frequencies uniquely determine
two free parameters, $\delta$ and $B_{LS}/B_{SS}$:
\begin{equation}
\frac{\nu_{j}}{\nu_{m}}
\simeq\frac{3}{2}\frac{B_{LS}}{B_{SS}}\,\delta,
\qquad
\frac{F(\nu_{j})}{F(\nu_{m})}\simeq\delta^2,
\end{equation}
which offers a unique diagnostic of GRB shocks.

\section{Conclusions}

We reviewed several models which has been proposed in order 
to resolve the puzzle of LoD-violating
and BPL gamma-ray bursts. It seems that the jitter model is the most
promising one, because it readily follows from the collisionless shock 
physics and results in minimal changes in the standard 
optically thin synchrotron shock model.







\end{document}